\documentclass[a4paper,11pt,twoside,openright]{article}
\usepackage{graphicx}
\usepackage{pdflscape}
\usepackage[square,sort,comma,numbers]{natbib}  
\usepackage{authblk}

\renewcommand{\rmdefault}{cmss} %helvetica
\fontfamily{\rmdefault}
\selectfont

\begin{document}

\title{Interests of a new lunar laser instrumentation on the ESO NTT Telescope}

 \author[1,2]{A. Fienga}
 \author[1]{C. Courde}
 \author[1]{J.M. Torre}
 \author[2]{H. Manche} 
 \author[3]{T. Murphy}
 \author[4]{J. Mueller}
 \author[2]{J. Laskar}
 \author[5]{S. Bouquillon}
  \author[4]{L. Biskupek}
   \author[4]{F. Hofmann}
    \author[5]{N. Capitaine }
 \author[2]{N. Rambaux}
 %\and A. Verma\inst{2}
 %\and P. Kuchynka\inst{1,3}

\affil[1]{AstroG\'eo, G\'eoazur-CNRS UMR7329, Observatoire de la C™te d'Azur,  Valbonne, France}
\affil[2]{Astronomie et Syst\`emes Dynamiques, IMCCE-CNRS UMR8028, Paris, France}
\affil[3]{Department of Physics, University of California, San Diego, La Jolla, CA, USA}
\affil[4]{Institut fur Erdmessung, Leibniz UniversitŠt Hannover, Germany}
\affil[5]{Observatoire de Paris, SYRTE CNRS-UMR 8630, Paris, France}

%\email{agnes.fienga@obs-besancon.fr}
%\renewcommand\Authands{ and }

% \offprints{A. Fienga}
%\institute{Astronomie et Syst\`emes Dynamiques,
% IMCCE-CNRS UMR8028,
% 77 Av. Denfert-Rochereau, 75014 Paris, France
% \and
% Observatoire de Besan\c con, CNRS UMR6213,
% 41bis Av. de l'Observatoire, 25000 Besan\c con, France
% }

 %\offprints{A. Fienga, agnes.fienga@obs-besancon.fr}

 \date{}

% \titlerunning{ INPOP10}
% \authorrunning{Fienga et al}

%  \abstract{
 
% \keywords{celestial mechanics - ephemerides}
% }
  \maketitle

%%-----------------------------------------------------------------
%%        The abstract
%% 
%%  Warning!  within the abstract:
%%  - do not use macros. 
%%  - do not use commands like: \cite, \citet, \citep ... etc.

\begin{abstract}
We analyze the impact of the installation of a lunar laser ranging device on the NTT 3.6m ESO telescope. With such an instrument, the scientific communities of fundamental physics and solar system formation will highly benefit of the only LLR station in the Southern Hemisphere. The quality of the NTT 3.6 meter telescope will also greatly  complement the LLR 3.5 meter Apache Point telescope (3.5 m) instrument in the Northern Hemisphere (USA) which is the best instrument for tracking the Moon since 2006. 
Finally, we also consider the technical characteristics of such installation including the observational constraints.
\end{abstract}

%% Insert the keywords (to appear in the ADS indexing)
%% Keywords must be separated by a comma
%\begin{keywords}
%Planetary ephemerides, numerical integration, space missions, tests of general relativity, asteroid masses
%\end{keywords}

%%-----------------------------------------------------------------

\section{Scientific rational: open questions addressed to Lunar Laser technology}
\subsection{Dark Matter and dark energy in the solar system and beyond}

The year 2015 will mark the 100th anniversary of General Relativity Theory (GRT). Up to now, GRT successfully described all available observations and no clear observational evidence against General Relativity was identified. However, the discovery of Dark Energy that challenges GRT as a complete model for the macroscopic universe and the continuing failure to merge GRT and quantum physics indicate that new physical ideas should be searched for. To streamline this search it is indispensable to test GRT in all accessible regimes and to highest possible accuracy.  Furthermore, the concordance of astrophysical measurements in the last 15 yearsÑthe anisotropy scale of the cosmic microwave background, the distance measurements of type Ia supernovae, the gravitational behaviors of galactic superclusters, and the power spectrum of large-scale structure Ñpoint to the surprising conclusion that the expansion of the universe is accelerating, implying some form of a fundamentally new gravitational phenomenon. The cosmological acceleration could be due to a scalar field that produces effects similar to those associated with the Ôcosmological constant,Õ originally introduced into the relativistic field equations by Einstein. 
Violations of the Equivalence Principle are predicted by a number of modifications of GRT aimed to suggest a solution for the problem of Dark Energy and/or to merge GRT with quantum physics (\cite{1994GReGr..26.1171D}, \cite{2010PhRvD..82h4033D}, \cite{2012CQGra..29r4001D}). The Universality of Free Fall (UFF), an important part of the Equivalence Principle, is currently tested at a level of about 10-13 with torsion balances (\cite{2003ARNPS..53...77A}) and the LLR (\cite{2012LPI....43.2230W}, \cite{2012AGUFM.P24B..02M}). A scalar field would also likely couple to the gravitational field in such a way as to produce a departure from the EP, and would introduce time variations in the fundamental coupling constants of nature. EP and time variations of G tests therefore have discovery potential with a very broad reach, and in fact provide some of the most sensitive low-energy probes for new physics. In light of recent discoveries, it is important that scientific inquiry is not restricted to current theoretical expectations, but rather that every available avenue for testing the nature of gravity is examined.
Some other formalisms often used to test gravity in the solar system and to solve some questions raised by the Dark Matter and the expending universe can also be tested with the LLR measurementsÊ: the modification of the inverse square law of gravity (\cite{2011RMxAC..40...11F}), additional force  represented by Yukawa-type expression (\cite{2003ARNPS..53...77A}, \cite{2011Icar..211..401K}).  Table 1 gives an overview of the performances obtained in the solar system for testing these formalims.

Measurement of the precession rate can also probe a recent idea (called DGP gravity) in which the accelerated expansion of the universe arises not from a non-zero cosmological constant but rather from a long-range modification of the gravitational coupling, brought about by higher-dimensional effects. Even though the lunar orbit is far smaller than the Gigaparsec length-scale characteristic of the anomalous coupling, there would be a measurable signature of this new physics, manifesting itself as an anomalous precession rate at about 5 $\mu$arcsec.yr$^{-1}$, roughly a factor of 10 below current LLR limits, and potentially reachable by millimeter quality LLR.
Tests of GRT remain to be very important tool to streamline the theoretical development. While a number of space missions are planned to improve these tests (MICROSCOPE to test the UFF with the level of 10-15, Gaia and BepiColombo to provide a number of high accuracy tests of GRT, EUCLID to study the distribution of Dark Matter in our Galaxy and the Universe, etc.), the instrumentation proposed here will lead to study the solar system dynamics for aiming at a set of advanced GRT tests that are complementary to the planned space-mission tests.
Finally, direct measurement of Dark matter in the solar system is also proposed by authors with the detection of its gravitational influence on the most accurately measured quantity in the solar system, the Earth-Moon distances (\cite{lrr-2010-7}). 

\begin{figure}
\begin{center}\includegraphics[width=14cm]{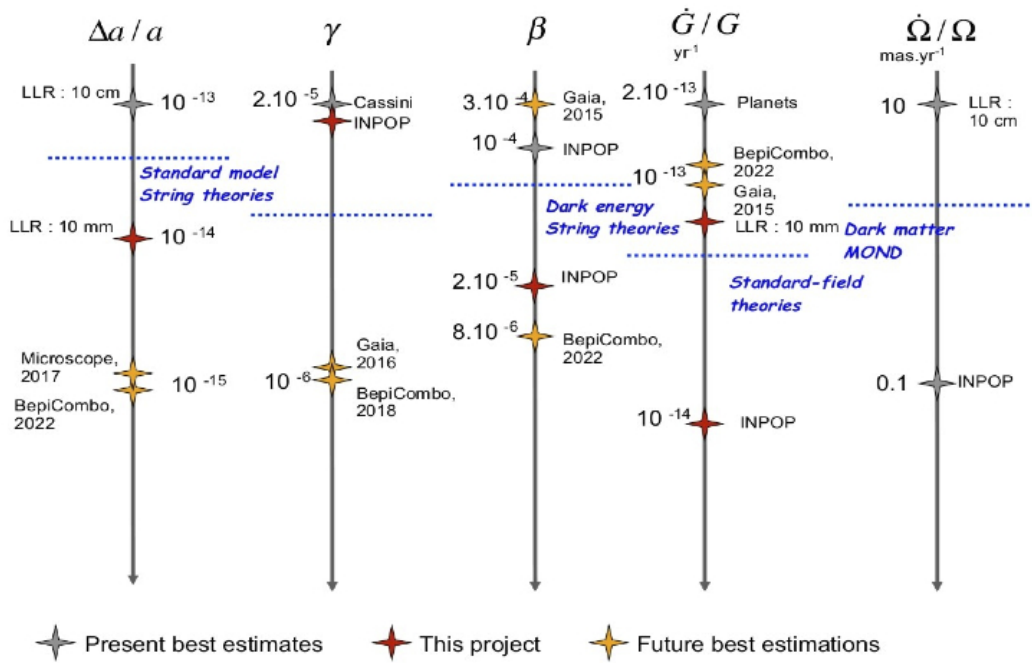}\end{center}
\caption{Present and expected results of gravity tests realized in the solar system}
\end{figure}

\subsection{Solar system and Moon formation}
The origin of the Earth-Moon system is currently being debated, the lunar bulk composition of
important refractory elements are unknown to within a factor of two, the origin of magnetic
anomalies are highly contested, and the size, composition, and state of the core are largely
unconstrained.  The Moon is the only terrestrial object for which we have samples from known locations, geophysical data from dedicated stations on the surface, and observations from field
geologists. From these data, the origin of the Moon from a giant impact with the early Earth,
the existence of a globe-encircling magma ocean that formed an ancient primary crust, the
existence of distinct geologic terranes, and a three-billion year record of volcanic activity
have been elucidated.
 However, because of the limited extent of the late 1960s digital technology, and the unfortunate placement of these stations near the boundary of the two most prominent geologic terranes, the interior structure and early geologic evolution of the Moon remains elusive.
Recent space missions such as LRO and GRAIL have opened new doors for the Moon comprehension but mainly for short periodic effects (high level gravitational field and topography) nut not for long term dissipative mechanisms (\cite{2012LPI....43.2230W}).
Nowadays, the rotational motion of the Moon is measured at the milli-second levels thanks to the LLR measurement of the round-trip travel time between an observatory on the Earth and one of the five corner cube retroreflector arrays on the Moon. The LLR data processing is a very sophisticated and challenging task with about 200 parameters depending the rotational and orbital models. The unprecedented accuracy in modeling the MoonÕs Dynamics at the centimetre and milli-arcsecond level is the result of recent developments in the laser station (OCA, APOLLO) and in the data processing.
Because of its long term coverage  (more than 40 years) of the Moon orbital and rotation states, the analysis of LLR data provides today informations on the lunar interior and solid-body tides, indicating that the lunar core is liquid (\cite{2001JGR...10627933W}, \cite{2004PhRvL..93z1101W}, \cite{2007GeoRL..34.3202W}) which has been confirmed by seismological studies (\cite{2011AGUFM.P31F..01W}). 
 In addition, the lunar rotational variations have strong sensitivity to moments of inertia and gravity field. The contributions to observations from tidal variations are sensitive to the interior structure of the Moon, its physical properties, and the energy dissipation inside the Moon are now at detectable level. Table 1 gives a global overview of the open questions in lunar physics and formation. 
 
 \begin{landscape}
 \begin{table}
 \caption{Open questions in lunar physics based on (\cite{2009IJMPD..18.1129W})}
 \begin{tabular}{c c c c}
 \hline
 Lunar core physical & Does the core is liquidÊ? &  Needs long time span & Core Moment of inertia \\
 characteristics (density, radius) &What sizeÊ? Is it uniformÊ? &of high accurate range data & Depending of the modele \\
& & & C/MR2 ratio\\
\hline
Constraints between the & Does a mantel existsÊ? What & Combination of long time & Full Moon moment of inertia \\
lunar core and the rest of & are the interactions between the & span of high accurate & \\
the body (mantel, crust & lunar mantel, crust and coreÊ? & range data but also s/c &  \\
etc...) & & tracking data & \\
\hline
Fluid-core/solid-mantle & Confirmation of a  fluid coreÊ? & Combination of long time & CMB oblateness \\
boundary (CMB) flattening & What is its viscosityÊ? What is & span of high accurate & and fluid core moment of \\
and dissipation & the mantel roughnessÊ?  & range data but also s/c & inertia, depending the model \\
& & tracking data & and energy dissipation \\
\hline
Inner Core & Does a solid inner core existÊ? & & Fluctuation of the 3 axis of \\
& What interaction with the mantelÊ? & & rotation of the Moon, resonances \\
\hline 
Elastic tides & Moon elastic properties  & New Lunar reflector locations & Love numbers k2 and l2 \\
\hline
Tidal dissipation & Value of the Q parameterÊ? & New Lunar reflector locations and & Q obtained from Moon \\
& Give constraints on core physics & long time span of high & libration and $\tau$ parameters \\
& & accurate range data & \\
\hline
Free libration & What mechanisms exist in the & future LLR measurements & Mode of rotations like \\
& Moon interiorÊ? What are  & & wobble mode of the lunar pole \\
& the interactions between & & \\
& the different layersÊ? & & \\
\hline
Orbital evolution & How the orbits of the Moon  & Detection of Tidal acceleration \\
& and the Earth have changed  & & Moon orbit with dissipation \\
& in the pastÊ? How they formedÊ? & & sources for the Moon and the Earth \\
\hline
 \end{tabular}
 \end{table}
 \end{landscape}
 
 \subsection{Earth rotation}
Lunar Laser Ranging (LLR), which has been carried out for more than 40 years, is used to determine many parameters within the Earth-Moon system (\cite{2013EGUGA..15.6131B}). LLR contributes to the determination of Earth orientation parameters (EOP) such as nutation, precession (including relativistic precession), polar motion, and UT1. The corresponding LLR EOP series is three decades long and contribute among others, to the determination of long-term nutation parameters, where again the stable, highly accurate orbit and the lack of non-conservative forces from atmosphere (which affect satellite orbits substantially) is very convenient. Additionally UT0 and VOL (variation of latitude) values are computed (e.g. \cite{1985JGR....90.9353D}), which stabilize the combined EOP series, especially in the 1970s when no good data from other space geodetic techniques were available. The precession rate is another example in this respect. The present accuracy of the long term nutation coefficients and precession rate fits well with the VLBI solutions (within the present error bars), see (e.g. \cite{2005LPI....36.1503W}). LLR can also be used for the realization of both the terrestrial and selenocentric reference frames. The realization of a dynamically defined inertial reference frame, in contrast to the kinematically realized frame of VLBI, offers new possibilities for mutual cross-checking and confirmation.

\subsection{New open activities and preparation of space missions}

The Moon physics cover a wide field of possible studies and ground based support for space lunar explorations or experiments could be done with the proposed facility. 
Furthermore, in the NASA Decadal survey, the Moon exploration is well positioned for selection for a medium mission launched in the next decade. Of particular interest are the sample return mission for South Pole Basin (Moonrise, \cite{2013LPICo1748.7034A}) and the Lunar Geophysical Network mission (LGN, \cite{2013LPICo1769.6017S}). Innovative proposals for installing new reflectors at the surface of the Moon in order to reach sub-millimeter level accuracy in the measurement of LLR distances were also emitted (\cite{2009arXiv0909.1509B}, \cite{2009IJMPD..18.1129W}, \cite{2013EGUGA..15.4330H}).  Such projects will need strong ground based support especially in the Southern Hemisphere where there is a critical lack of tracking stations in radio (DSN network) but also in optical wavelength (LLR network) reaching interplanetary distances. Synergies between LLR tracking and lunar seismology (\cite{2009AGUFM.P23C1279N}), spacecraft gravity field determinations and spacecraft laser tracking (\cite{2013JGRE..118.1415K}), magnetic field  explorations (\cite{2009LPICo1483...56H}), studies of volatiles at the surface of the Moon  are also some possible examples of ground base  LLR support required by space missions. 

\section{Why a new LLR instrumentation on the ESO NTT ?}

Nowadays only three laser stations in the Northern hemisphere (the Apache Point telescope (APOLLO, USA) since 2006, Mac Donald Laser Ranging Station (MLRS, USA) and Observatoire de la C™te d'Azur (OCA, France))  are able to operate regular interplanetary laser ranging towards the Moon surface and the five usable retroreflectors on its near side.  While a single Earth ranging station may in principle range to any of the these  retroreflectors  from any longitude during the course of an observing day, these observations are nearly the same in latitude with respect to the Earth-Moon line, weakening the geometric strength of the observations. In the present time, there is no stations capable of ranging to the Moon in the Southern Hemisphere. Furthermore the frequency and quality of observations varies greatly with the facility and power of the laser employed.  During the past 40 years, the OCA team have a unique background in Lunar Laser Ranging with the OCA laser station,  providing 60 \% of the LLR data and developing upgraded instrumentations. The APOLLO (the Apache Point Observatory Lunar Laser-ranging Operation) station has started its LLR activity in 2006 providing the most accurate and numerous (50\% of the global LLR data since 2006 are APOLLO's, OCA 30\% and MLRS 14\%) observations thanks to its high quality electronic devices and the 3.5m aperture telescope localized in altitude in the Northern Hemisphere. Based on the aperture similarities and observational strategies (few hours during the night), a LLR instrumentation on the NTT (in the following called SHELLI for Southern Hemisphere Lunar Laser Instrument) will be seen in the next simulations as a twin of APOLLO localized at La Silla in terms of quality and regularity of the produced data.

\subsection{Impact of a Southern Hemisphere station}

Improvements in the geometric coverage, both on Earth and on the Moon, will have a direct impact on the science gained through LLR. Studies of the structure and composition of the interior require measurements of the lunar librations, while tests of GR require the positions and velocities of the lunar centre of mass. In all, six degrees of freedom are required to constrain the geometry of the Earth-Moon system (in addition to Earth orientation). A single ranging station and reflector is insufficient to accurately determine all six degrees of freedom, even given the rotation of the Earth with respect to the Moon. The addition of one or more reflectors and one or more additional ranging stations in the EarthÕs Southern Hemisphere would strengthen the geometric coverage and increase the sensitivity to lunar motion by as much as a factor of 4 in some degrees of freedom at the same level of ranging precision (Hoffmann et al. 2013). 
We operated computations considering simulated observations obtained by a LLR station located at La Silla. In using the INPOP planetary and Moon ephemerides (\cite{Fienga2011}), we have simulated the construction of new orbital and rotational ephemerides of the Moon  including 50\% of Northern Hemisphere observations and 50\% of  simulated Southern Hemisphere data. This percentage is realistic as it corresponds to the percentage of data obtained by the APOLLO instrument during the past 6 years. Based on the weather conditions at La Silla, one can expect to reach this level of sampling after 4 years of SHELLI exploitation. In considering an uncertainty for the Southern Hemisphere data equivalent to regular (OCA and MLRS) LLR observations (with an accuracy of about 2cm), we have obtained  improvements in the determination of dynamical parameters (reflector positions, geocentric Moon positions and velocities) but also for internal parameters as presented in Table 1. These results demonstrate the interest of developing a Southern Hemisphere LLR station especially for dynamical parameters, important for testing gravity and fundamental physics but also for a better knowledge of the size and density of the inner core. 

As it was stated previously, for now there is no Lunar Laser Ranging station in the Southern Hemisphere. Projects in South Africa and Australia are no longer in question and discussions are nowadays lead in China, Japan and India for installing such facility. 

\begin{table}
\caption{Improvements  in Moon dynamical and internal parameter estimations obtained by including simulated Southern Hemisphere observations (with a simulated accuracy of 2 cm) and 3.6m  Southern Hemisphere observations (with a simulated accuracy of 0.5 cm).}
\begin{tabular}{l c c l}
\hline
Parameter & Southern Hemisphere & 3.6 m Southern station & What for ? \\
\hline
Geocentric position  & 10 to 30 \% & 25 to 40 \% & Dynamics, \\
and velocity of the Moon & & & Tests of GR \\
Gravity Field coefficients & 2 to 30\% & 15 to 30\% & Dynamics,  \\
& & & Libration \\
C/MR2 ratio & 15\%& 25\% & Inner core size \\
& & & and density \\
Love Numbers& 10\% & 25\% & Moon Elasticity \\
$\tau$ parameters & 5\% & 25\% &  Q of dissipation \\
Mass of the Moon-Earth system & 10\% & 30\% & Dynamics,  \\
& & &  Tests of Gravity \\
Positions of the reflectors & 5\% & 20\% & Dynamics, Libration \\
 at the Moon surface & & & , Tests of GR\\
 \hline
\end{tabular}
\end{table}

\subsection{Impact of a 3.6 m Southern Hemisphere station}
The first LLR measurements had a precision of about 20 cm. Over the past 35 years, the precision has increased only by a factor of 10. The  APOLLO instrument has gained another factor of improvement, achieving the sub-centimeter level precision. Poor detection rates are a major limiting factor in past LLR. Not every laser pulse sent to the Moon results in a detected return photon, leading to poor measurement statistics. A classic Lunar Laser Station (e.g. MLRS) typically collects less than 100 photons per range measurement with a scatter of about 2 cm. The large collecting area of the APOLLO (3.5 m telescope) comparable to NTT and the efficient avalanche photodiode arrays used in the APOLLO instrument proposed to be installed at NTT reach thousands of detections (even multiple detections per pulse) leading to a potential statistical uncertainty of about 1 mm. The simulations presented in Table 2 illustrate the important improvement brought by a 3.6m technology in the Southern Hemisphere in considering a mean accuracy in the simulated observations of about 0.5 cm (to be compared to the 2 cm of the previous simulation). The uncertainties of the Moon dynamics are reduced by a factor 3 when the coefficients related to the dissipation and those of the elasticity of the Moon interior are obtained with an accuracy improved by a factor 2.5. Some coefficients of the Moon gravity field (C30, C31, S31, C33, S33) face an improvement of one order of magnitude compared to what could be obtain by a reduced aperture telescope in the Southern Hemisphere. These results confirm the fundamental impact of the installation of a highly efficient Lunar Laser instrumentation in the Southern Hemisphere for testing General Relativity and Dark matter/Dark energy physics, for a better knowledge of the dissipation mechanism between the lunar core and mantle and a better understanding of the inner core physical characteristics.

\subsection{Observational strategies and potential users}
2 strategies can be proposed to the users:
\begin{itemize}
\item For teams interesting in improving the Moon dynamics and testing general relativity, 1-hour session of observations is necessary at the starting of the daily Moon observational period that could start during the day and a 1-hour session at the end of the daily session. Regular observations obtained during 1-hour sessions are important for improving Moon dynamics and inner physics. It has to be stressed that these short sessions can happen during the day and are suspended twice per months, during full and new Moon periods. About 50\% of the sessions will be done by day. Constraints related to the NTT such as limitation in zenith distances are to be taken into account but will not affect greatly the regularity of the obtained samples as the NTT constraints are quite similar to other LLR stations. 
\item For teams interesting in Earth rotation, a session of several hours for the obtention of the longest data arc is necessary. This type of observational set can be asked by users during a limited period of time for the improvement of parameters related to Earth rotation and more specifically nutation. 
\end{itemize}
The scientific community interested in such instrument is important as the Moon is a critical tool for a lot of different fields of research and activities. It gathered an important number of specialists over the word covering different fields of research from fundamental physics to the scenarii of formation of the solar system but also Earth rotation and  reference frame definitions.

\section{Description of the technical proposal}

Lunar Laser Ranging is based on a network of laser stations distributed ideally around the earth. These stations use the time of flight of laser pulses to determine distance between ground instrument and retro-reflector installed on the moon. 
The advantage of Laser Ranging lies in the simplicity of the measurement principle and the use of completely passive space segment (low cost). Its accuracy is mainly based on very accurate time frequency standards.  One of the first general requirements for LLR station site is to optimize the place in respect with the cloud coverage. In an area  some specific zones can have often fog or mist and in the same time some kilometers from there, the sky can be clear. These meteorological characteristics affect dramatically the link budget and so the number of measurements. The sky quality at the La Silla Observatory offers the guarantee to have the best conditions of measurement. 
Based on the experiences of OCA and APOLLO teams, we proposed to install a Lunar Laser device on the Nasmyth focus of NTT following the three requirements:
\begin{itemize}
\item a telescope able to follow the moon with a tracking accuracy below 1 arcsecond
\item a high energy, green, pulsed laser 
\item a detection channel able to detect and to date single-photon events with sub-picosecond precision
\end{itemize}
Compared to the OCA telescope with its primary mirror of 1.5 meters in diameter, the number of photon received with the NTT will be at least 8 times greater. Moreover the NTT is 1000 meters higher in altitude than the OCA station. It increases also the link budget between the received and emitted photon. 
Laser and detection channels can be implemented together in the Nasmyth focus of the NTT by duplicating the OCA compact instrument developed since many years. We propose to design and construct an optical table on which all the requested instrumentation  will be set up. The equipments will be buy or developed and tested at OCA. The operation of the optical bench will be validate on OCA telescope before to be sent for La Silla. Finally, the optical bench will be implemented and validate on the NTT. 
\begin{figure}
\caption{Optical Bench to be installed at the NTT Nasmyth focus.}
\begin{center}\includegraphics[width=12cm]{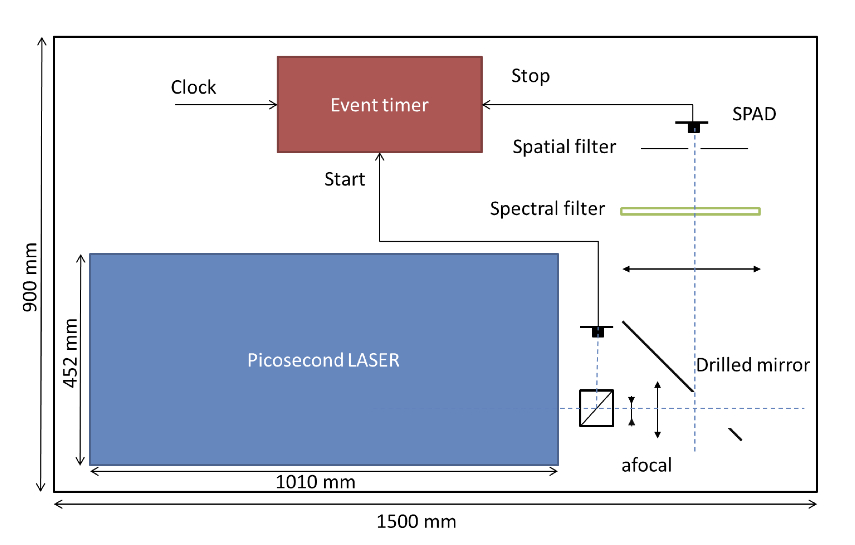}\end{center}
\end{figure}
The figure 2 gives an idea of what will look like the optical bench implemented in the Nasmyth focus. The size of the optical bench is 900 x 1500 x 200 mm. For the uplink, a high-energy mode-locked Nd:YAG picosecond  laser will be buy. For LLR, we need of less 150 mJ/pulse at 532 nm with a pulse width around 100 ps. This kind of laser is commercially available. Divergence of the laser beam will be managed by an afocal (adjusting finely the distance between two lens). A small part of the laser is sent to a high speed photodetector linked to an event timer to determine the pulse departure.  
For the downlink, a scientific camera is needed to point the telescope on the retro-reflector location with accuracy better than 1 arcsecond. We will use the guiding camera already available at NTT. To detect the laser echoes, we use a single photon avalanche diode (SPAD) in Geiger mode. Due to its very high sensitivity, the SPAD is triggered a short time before the photon arrival. To reduce the noise, the SPAD is placed behind a spectral and a spatial filter. The detection signal is sent to the event timer to determine the pulse arrival. The precision of the measurement depends on the clock performances used by the event timer. Atomic-clocks (Cs or Rubidium) locked on GPS time have the performances required for this purpose. They are very versatile system. The main constraint is to install the GPS antenna outdoor. Then the clock is completely automated and can be implemented in a rack. On the OCA station, the switch between the up and the down link is done by a rotating mirror. We are testing a second technique completely passive named aperture sharing: the up link is emitted on a small surface of the primary mirror. The down link uses the remaining surface (as illustrated on the previous figure).  This new development on OCA telescope could therefore be implemented on NTT.  

The  figure 3  presents an overview of the instrumentation installed at La Silla. Management of laser and photodetection will be controlled remotely. Dedicated software will be delivered with the optical bench. 
The pass predictions will be calculated by our computer server and sent to the control station of the NTT. The telescope will follow the pass prediction and the monitoring work will consist of making pointing correction in order to obtain laser echoes. 
A weather station will be installed outdoor. Temperature, pressure and humidity measurements are needed for the measurement accuracy and the correction of atmospheric effects. 
We insist on the fact there is no modification telescope in our technical proposal. All the instrumentation will be installed at the Nasmyth focus or outdoor. Moreover, there is no innovative technical development which could penalize the project. Innovation held in our ability to quickly install this kind of metrology on the NTT. 

\begin{figure}
\caption{Instruments to be installed at the NTT telescope}
\begin{center}\includegraphics[width=10cm]{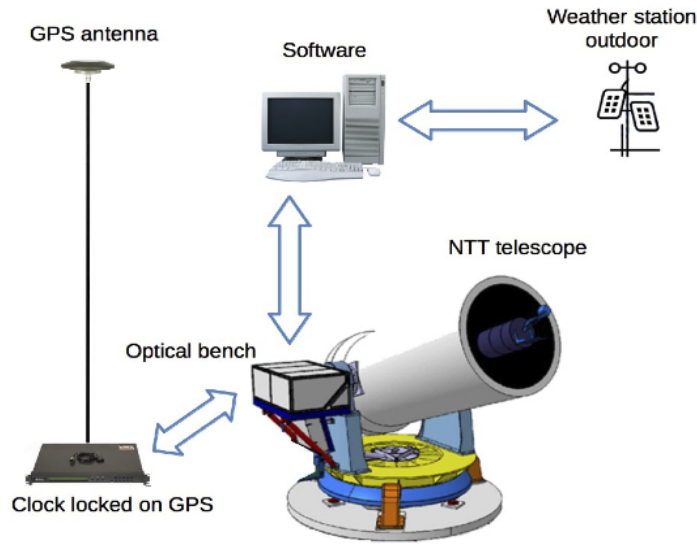}\end{center}
\end{figure}

\section{Conclusions}

We give in this paper a description of a possible laser instrument to be installed at the ESO NTT telescope.
We stress the  scientific interest of such an instrument not only for a better understanding of the lunar inner physics but also for testing gravity in the frame of alternative theories including Dark matter density in the solar system.

\section{References}

\bibliography{biblio_hdr}{}
\bibliographystyle{plain}

\end{document}